\documentclass[english]{article}
\usepackage[T1]{fontenc}
\usepackage[latin9]{inputenc}
\usepackage{geometry}
\usepackage{amssymb}
\usepackage{graphicx}
\usepackage{babel}

\usepackage{color}

\def\D{
{D\mkern-2mu\llap{{\raise+0.5pt\hbox{\big/}}}\mkern+2mu}
}
\def\nablaslash{
{\nabla\mkern-2mu\llap{{\raise+0.5pt\hbox{\big/}}}\mkern+2mu}
}
\def\be{\begin{equation}}
\def\ee{\end{equation}}
\def\semidirectproduct{
{\ooalign
{\hfil\raise.07ex\hbox{s}\hfil\crcr\mathhexbox20D}}}

\begin{document}

\title{Non relativistic SUSY in variants of the planar Lévy-Leblond equation}

\author{L\'aszl\'o Palla\footnote{email: palla@ludens.elte.hu}}

\maketitle

\begin{center}
\emph{Institute for Theoretical Physics, Roland E\"otv\"os University,}\\
\emph{ 1117 Budapest, Pázmány Péter sétány 1/A, Hungary}
\end{center}

\begin{abstract}
  An $N=2$ SUSY extension of the Schrödinger symmetry is shown to exist in
  the solution space of 
  the free planar L\'evy-Leblond equation, an $N=1$ part of which survives for
  the gauged version of the equation and also when it is coupled to Chern-Simons  theory.
\end{abstract}

\section{Introduction}

A natural way to obtain models with non relativistic SUSY is to consider the
non relativistic limit of relativistic, supersymmetric theories
\cite{Leblanc:1992ns}. Here we
are interested rather in extending the combination of Galilean and
non relativistic conformal symmetries \cite{Jackiw:1972nc} 
(called Schrödinger symmetry) by
anticommuting generators in models/equations containing no apparent
supersymmetry or superpartners. SUSY extensions of the Schrödinger symmetry
were considered previously \cite{Beckers:1986ab} \cite{Jackiw:1990su} 
\cite{Horvathy:1993fd} \cite{Duval:1993hs} \cite{Duval:1995dq}.
In \cite{Duval:1993hs} it is shown that in $d>2$ space dimensions
Schrödinger symmetry admits a unique $N=1$ extension, while in $d=2$ there are
two such extensions that combine into an $N=2$ extension.
In \cite{Duval:1995dq} a physical
realization of this special 
extension is described on the example of a spin $1/2$
particle moving in the field of a magnetic vortex.

In this paper we look for realization of supersymmetry in the solution space
of certain variants of the {\bf planar} Lévy-Leblond equation namely in the
free and in the gauged equations and when it is coupled to Chern Simons theory.
The $3+1$ dimensional Lévy-Leblond equation \cite{Levyleblond:1967ll} 
may be thought of as a
non relativistic Dirac equation: it is a first order differential equation for
a spin $1/2$ particle, the square of which gives the Pauli equation. For the
$3+1$ dimensional free Lévy-Leblond equation an $N=1$ extension of the
Schrödinger symmetry is described in \cite{Aizawa:2016vmo}.
Since the Schrödinger symmetry for
these three variants of the  planar Lévy-Leblond equation is shown
in \cite{Duval:1995fa} 
we look here for anticommuting generators (that also anticommute with the
Lévy-Leblond differential operator) that extend the symmetry. 

The motivation to study these planar systems comes not only from mathematics but
also from physics as $2+1$ dimensional Chern Simons electrodynamics is
generally thought to provide a viable alternative to describe interesting
physical phenomena like high $T_c$ superconductivity \cite{Chen:1989ab} 
or the quantized Hall effect \cite{Zhang:1988wy}.

 We investigate 
this possible non relativistic supersymmetry in variants of the
 planar L\'evy-Leblond equation (LLE) in a Kaluza -Klein type framework.
The main idea is that non relativistic $2+1$ dimensional space time $R$ may be
viewed as the quotient of a $3+1$ dimensional Lorentzian manifold $(M,g)$ by
the integral curves of a covariantly constant light-like vector $\xi$ (such a 
manifold is called a Bargman space) \cite{Duval:1985bs}. In this
framework the non relativistic symmetries are the higher dimensional ones
leaving
$\xi$ invariant. An adapted coordinate system on $M$ is given by
$(t,x^j,s)$ ($j=1,2$), where $\xi\equiv\partial_s$, and
$(t,x^j)$ are coordinates on
$R$, such that $(x^1,x^2)$ give the positions and $t$ is non relativistic
absolute time. In this paper we consider the case when $M$ is flat Minkowski
space with metric $ds^2=\sum (dx^i)^2+2dtds $.

The paper is organized as follows: In sect.2 we derive the $N=2$ extension of
the Schrödinger symmetry for the free planar LLE, sect.3 deals with the gauged
LLE, while in sect.4 we investigate the case when the LLE is coupled to
Chern-Simons theory. In each of these three sections we first review what is
known about the Schrödinger symmetry in that particular case before we look
into the
supersymmetric extension. We make our conclusions in sect.5, which is followed
by three appendices. 

\section{Search for non relativistic SUSY in the planar L\'evy-Leblond equations}

\subsection{Free LLE and its bosonic symmetries}

First we consider the {\bf free} LLE; 
we obtain them from the free massless Dirac equation on $4$d Minkowski space 
$$
\nablaslash\psi =0\,.
$$
Using the Dirac matrices (satisfying $\{\gamma^\mu,\gamma^\nu\}=-2g^{\mu\nu}$) 
\be
\gamma^t=\pmatrix{0&0\cr1&0},
\qquad
\gamma^i=\pmatrix{-i\sigma^i &0\cr0&i\sigma^i},\quad i=1,2
\qquad
\gamma^s=\pmatrix{0&-2\cr0&0},\label{gm}
\ee
the equivariance condition $\nabla_\xi \psi=im\psi $, ($\xi\equiv\partial_s$) 
and the Ansatz
$\psi=e^{ims}\pmatrix{\Phi\cr\chi}$ (where $\Phi$ and $\chi$ depend on $t$ and
$x^j$ only) 
in the Dirac equation we find
\begin{equation}
  \pmatrix{-i\sigma^j\partial_j&-2im\cr\partial_t&i\sigma^j\partial_j}
  \pmatrix{\Phi\cr\chi}=0\,.
  \label{freeLL}
\end{equation}
The 4d chirality matrix
$$
\Gamma=-\frac{\sqrt{-g}}{4!}\,\epsilon_{\mu\nu\rho\sigma}
\gamma^\mu\gamma^\nu\gamma^\rho\gamma^\sigma,\qquad\quad  
\Gamma=\pmatrix{-i\sigma^3&0\cr 0&i\sigma^3}
$$ splits
eq.(\ref{freeLL}) into two uncoupled equations for the chiral components
$\Gamma\psi_\epsilon=-i\epsilon\psi_\epsilon$, $\epsilon=\pm $\footnote{As shown
  in \cite{Duval:1995fa} the
$\psi_\pm$ describe the spin $1/2$ (spin $-1/2$) representations of the $2+1$
  dimensional Schr\"odinger group.}. As discussed in \cite{Duval:1995fa} the
independent
(two component) equations for $\psi_\pm$
$$
\psi_+=e^{ims}\pmatrix{\phi_+\cr0\hfill\cr0\hfill\cr\chi_+},\qquad
\pmatrix{-i(\partial_1+i\partial_2)&-2im\cr\partial_t&i(\partial_1-i\partial_2)},
\pmatrix{\phi_+\cr\chi_+}=0,
$$
$$
\psi_-=e^{ims}\pmatrix{0\hfill\cr\phi_-\cr\chi_-\cr0\hfill},\qquad
\pmatrix{-i(\partial_1-i\partial_2)&-2im\cr\partial_t&i(\partial_1+i\partial_2)},
\pmatrix{\phi_-\cr\chi_-}=0, 
$$
are the two possible (free) LL
equations in two spatial dimensions. (The existence of two LLE is a special
 property of two spatial dimensions). 
Nevertheless, for reasons becoming clear below, (and to describe them
simultaneously)   
we keep the 4d matrix form
of the LLE even for these chiral components. Note that the two component
$\Phi=\pmatrix{\phi_+\cr\phi_-}$ and $\chi=\pmatrix{\chi_-\cr\chi_+}$ are not
chiral, but are composed in a particular way of the components of $\psi_\pm$. 
Since
$\nablaslash\nablaslash =(-2im\partial_t -\partial_k\partial_k)\mathrm{\bf 1}_4$,
the \lq\lq square'' of the LLE can be written as
\begin{equation}
i\partial_t\pmatrix{\Phi\cr\chi}=H\pmatrix{\Phi\cr\chi}, \qquad\qquad 
H=-\frac{1}{2m}\partial_k\partial_k\mathrm{\bf 1}_4\,.
\label{freeSch}
\end{equation}
This implies, that every component of every solution of any of the free LLEs  
automatically solves the free Schr\"odinger eq.

In \cite{Duval:1995fa} we showed that the $\xi(\equiv\partial_s)$ preserving conformal
transformations (besides preserving the equivariance and chirality conditions) 
are symmetries of the free LLE. This means that the operators
(${\cal B}$) implementing the infinitesimal transformations satisfy
\be
   [{\cal B},\nablaslash]=\Sigma_{\cal B}\nablaslash , \label{symeqb}
   \ee
   where $\Sigma_{\cal B}=-\frac{1}{4}\nabla_\mu X_{\cal B}^\mu$ with $X_{\cal B}^\mu\partial_\mu$ being the vector field describing the conformal transformation on
   Minkowski space.  
    For later reference the explicit forms of dilatation ($d$),
Galilean boost ($b^j$), expansion ($K$) and rotation ($J$) are listed here
as they follow from eq.(4.7) and (4.1) in \cite{Duval:1995fa}:
 $$
 d=\pmatrix{2t\partial_t+x^k\partial_k+1&0\cr 0&2t\partial_t+x^k\partial_k+2},
 \quad b^j=\pmatrix{t\partial_j-imx^j&0\cr\frac{i}{2}\sigma^j&t\partial_j-imx^j},
 $$
 $$
 K=\pmatrix{t^2\partial_t+tx^k\partial_k+t-\frac{im}{2}r^2&0\cr\frac{i}{2}\sigma^jx_j&t^2\partial_t+tx^k\partial_k+2t-\frac{im}{2}r^2},\qquad r^2=x^kx^k,
 $$
 $$
 J=-\pmatrix{x_1\partial_2-x_2\partial_1+\frac{i\sigma^3}{2}&0\cr0&x_1\partial_2-x_2\partial_1+\frac{i\sigma^3}{2}}.
$$
 (The time translation ($\sim\partial_t$) and spatial translation ($\sim\partial_j$) operators are trivial). All $\Sigma_{\cal B}$ vanish with the exception of
 $\Sigma_{d}=-\mathrm{1}_4$ and $\Sigma_{K}=-t\mathrm{1}_4$. The structure of
 the symmetry algebra (called planar Schrödinger algebra $sch(2)$) is the
 following: translations and boosts form a Heisenberg algebra $h(2)$ (
 $[i\partial_j,b^k]=m\delta^{jk}\mathrm{1}_4$), $(i\partial_t,d,K)$ form the
 $sl_2$ algebra of 
 nonrelativistic  conformal symmetry  \cite{Jackiw:1972nc},
 \begin{equation}
   [d,i\partial_t]=-2i\partial_t,\qquad
[i\partial_t,K]=id,\qquad 
 [d,K]=2K,
 \label{dhk}
 \end{equation}
 $J$ forms the $so(2)$ of planar
 rotations and they combine into $sch(2)$ as 
 $sch(2)=(sl_2\oplus so(2))\semidirectproduct
 h(2)$. We emphasize, that all generators commute with $\Gamma$, indicating
 that $\psi_+$ and $\psi_-$ span two different representations.
 
 Anticipating 
  the forthcoming question of the $sl_2$ subalgebra containing $H$, $d$ and
 $K$ we compute 
\be
 [d,H]=-2H. 
\label{T}
\ee

\subsection{Supercharge candidates and the concept of weak identification}

Now we look for the
``fermionic'' extensions of these bosonic symmetries, i.e. for ${\cal F}$-s
satisfying 
\begin{equation}
  \{{\cal F} ,\nablaslash\}=\Sigma_{\cal F}\nablaslash , \label{symeq}
\end{equation}
where $\Sigma_{\cal F}$ may depend on the coordinates, but cannot contain derivatives \cite{Aizawa:2016vmo}. We found the following trivial solutions:
$$
{\cal F} =\mathrm{\bf 1}_4,\quad\Sigma_{\mathrm{1}_4}=2\mathrm{\bf 1}_4;\qquad\quad
{\cal F} =\Gamma ,\quad\Sigma_\Gamma=0;
$$
together with two non-trivial ones:
$$
{\cal F} =\tilde{Q}=\frac{1}{\sqrt{2m}}
\pmatrix{-i\epsilon_{kl}\sigma^k\partial_l&0\cr 0&i\epsilon_{kl}\sigma^k\partial_l}=\frac{1}{\sqrt{2m}}\gamma^k\epsilon_{kl}\partial_l,\qquad\qquad
\Sigma_{\tilde{Q}}=0,
$$
and
\be
   {\cal F}=\Lambda=\pmatrix{0&\beta\cr\alpha\partial_t&0}=
   \alpha\gamma^t\partial_t-\frac{\beta}{2}\gamma^s,\qquad\qquad
\Sigma_\Lambda=0,\qquad{\rm if}\qquad \beta-2im\alpha=0.
\label{Lambdadef}
\ee
The first two transformations generate a (continuous) chiral rotation
$\exp(\alpha\Gamma)$ that multiplies $\psi_\pm$ by $\exp(\pm i\alpha)$ and are thus
not very interesting.

The interesting solutions are 
third one (which is the 4d version of
the ``twisted'' $Q_{\rm 2d}$ \cite{Jackiw:1990su} \cite{Duval:1995dq})
and which is normalized such that $\tilde{Q}\tilde{Q}=H$,
and the fourth one with square
 $\Lambda\Lambda=\alpha\beta\partial_t\mathrm{1}_4$ which is proportional to
$\partial_t$, just like in the $3+1$ dimensional case investigated
in \cite{Aizawa:2016vmo}. Both $\tilde{Q}$ and $\Lambda$ commute with
$\xi\equiv\partial_s$ ($[\tilde{Q},\xi]=0=[\Lambda,\xi]$), thus the symmetry
they generate descends to the planar equations.  A direct computation shows,
that
\be
\{\Gamma ,\tilde{Q}\}=0,\qquad \{\Gamma,\Lambda\}=0,\qquad
\{\Lambda,\tilde{Q}\}=0.
\label{lambdakuanti}
\ee
Therefore both $\tilde{Q}\psi_\epsilon$ and $\Lambda\psi_\epsilon$ 
have opposite chirality to $\psi_\epsilon$. (In $2+1$
dimensions this means that the $1/2$ and $-1/2$ spin representations are
changed into each other by $\Lambda$ and 
$\tilde{Q}$). Furthermore it also means, that if
$\psi_+$ solves its LL equation then $\tilde{Q}\psi_+$ (or $\Lambda\psi_+$) 
solves the equation for
$\psi_-$ rather than the one for $\psi_+$. Thus, strictly speaking, neither 
$\tilde{Q}$ nor $\Lambda$ is a symmetry of either the $\psi_+$ or the $\psi_-$
equations, only of the system in (\ref{freeLL}), and to represent these
operators we need
both the spin $1/2$ and $-1/2$ spinors.

$\tilde{Q}$ and $\Lambda$ commute with translations and are scalar under
rotation
$$
[i\partial_j,\Lambda]=0,\quad [J,\Lambda]=0,\qquad
[i\partial_j,\tilde{Q}]=0,\quad[J,\tilde{Q}]=0,
$$
and have the same dimensions
$$
[d,\Lambda]=-\Lambda,\qquad [d,\tilde{Q}]=-\tilde{Q}.
$$
Adding to this the fact that $\tilde{Q}\tilde{Q}=H$, and 
$\Lambda\Lambda=\alpha\beta\partial_t\mathrm{1}_4$ makes one wonder whether one
can use them as supercharge candidates in the sought of fermionic extension. For
a supercharge its square should be (proportional to) a bosonic generator; this
condition is met for $\Lambda$, but not - at least naively - for $\tilde{Q}$; as
the bosonic generators listed above contain only first order derivatives. Thus
it seems that $\tilde{Q}$ is eliminated. 

However, if we
choose $\alpha\beta=i$, which implies
\be
\alpha=\frac{1}{\sqrt{2m}},\quad 
\beta=i\sqrt{2m},\qquad{\rm and}\qquad 
\Lambda=\pmatrix{0&i\sqrt{2m}\cr\frac{\partial_t}{\sqrt{2m}}&0};
\label{alfabeta}
\ee
 then, for {\bf solutions} of LLE, we can write
 \begin{equation}
 \Lambda\Lambda=i\partial_t\mathrm{1}_4=H=\tilde{Q}\tilde{Q},
 \label{thid}
 \end{equation}
 in light of the Schr\"odinger eq., (\ref{freeSch}). Putting it differently,
 for solutions of LLE (``weakly''), we identify $i\partial_t$ and
 $-\frac{1}{2m}\partial_k\partial_k$. Of course we have to check whether in the 
 bosonic algebra one can consistently make this identification.  Since $H$
 trivially commutes with translations and rotation just like $i\partial_t$ and 
 a direct algebraic computation gives
 $$
 [i\partial_t,b^j]=i\partial_j,\qquad [H,b^j]=i\partial_j,
 $$
 we look whether $(H, d, K)$ also form (only \lq\lq weakly'' of course)
 an $sl_2$ algebra. In the light of eq.(\ref{dhk},\ref{T}) we must compute
 $[H,K]$ to
 check this. We find algebraically
 $$
 [H,K]=\pmatrix{2t(-\frac{1}{2m}\partial_k\partial_k)+ix^k\partial_k+i&0\cr
   -\frac{i}{2m}\sigma^k\partial_k&2t(-\frac{1}{2m}\partial_k\partial_k)+ix^k\partial_k+i}\,.
 $$
 Using the identification (\ref{thid}) this can be written as
 $$
 [H,K]=id+\frac{1}{2m}\pmatrix{0&0\cr-i\sigma^k\partial_k&-2mi}
 =id+\frac{1}{2m}\gamma^t\nablaslash.
 $$
 However, for solutions of the LLE, eq.(\ref{freeLL}), (i.e. \lq\lq weakly'')
 the second term vanishes. Therefore we can say that on the solution manifold
 of LLE $(H, d, K)$ also form an $sl_2$ algebra, thus the identification in 
 (\ref{thid}) works. One may wonder whether the weakly vanishing term appearing
 in $[H,K]$ preserves weakly also the Jacobi identity, i.e. 
 generates only further weakly vanishing terms when one forms the commutators
 $[{\cal B},[H,K]]$ for any ${\cal B}$ in the Jacobi identity. We investigate
 this question in Appendix A in a somewhat wider context, when we consider
 also the various weakly vanishing terms coming from the (anti)commutators of
 the forthcoming fermionic operators.

 \subsection{The fermionic extensions}

 One can obtain new fermionic symmetry 
 generators by commuting $\Lambda$ and $\tilde{Q}$ with the
 bosonic generators. We start with $\Lambda$, and since $\Lambda$ commutes with translations and rotation   
 consider first 
 $$
 [\Lambda,b^j]:=Z^j,\qquad j=1,2\qquad Z^j=\pmatrix{i\frac{\beta}{2}\sigma^j&0\cr\alpha\partial_j&-i\frac{\beta}{2}\sigma^j}=\frac{1}{\sqrt{2m}}\pmatrix{-m\sigma^j&0\cr\partial_j&m\sigma^j}.
 $$
 If $\Lambda$ anticommutes with $\nablaslash$, then,
 since $[b^j,\nablaslash ]=0$, $Z^j$ should also anticommute with $\nablaslash$,
 i.e. they also generate a fermionic symmetry. They have vanishing dimension
 and their anticommutator with $\Lambda$ is proportional to translation:
 \be
 [d,Z^j]=0,\qquad\qquad \{\Lambda,Z^j\}=i\partial_j\mathrm{1}_4.
 \label{Z1}
 \ee
 The anticommutator of the (fermionic) $Z^j$-s contains a central element
 \be
 \{Z^j,Z^k\}=m\delta^{jk}\mathrm{1}_4,
 \label{Z2}
 \ee
 showing they form a fermionic Heisenberg algebra.

 The second operator we introduce is the commutator of $\Lambda$ and
expansion:
$$
[\Lambda,K]:=\hat{S},\qquad\quad
\hat{S}=\frac{1}{\sqrt{2m}}\pmatrix{-m\sigma^jx_j&2im t\cr(t\partial_t+x^j\partial_j+1)&m\sigma^jx_j}.
$$
One can check, that $\{\hat{S},\nablaslash\}=0$, 
thus this operator also generates a fermionic symmetry. The dimension of
$\hat{S}$ follows from that of $\Lambda$ and $K$: $[d,\hat{S}]=\hat{S}$. 
Furthermore 
the ``square'' of this operator is proportional to expansion 
\be
\hat{S}\hat{S}=i K,
\label{SS}
\ee
showing that $\hat{S}$ may be thought of as a
conformal supercharge. At this point the construction of new fermionic
generators with the aid of $\Lambda$ comes to an end: 
we can not repeat this procedure starting with 
$Z^j$ or $\hat{S}$ since
$$
[Z^j,b^k]=0,\qquad [\hat{S},b^j]=0,\qquad [Z^j,K]=0,\qquad [\hat{S},K]=0.
$$

Next we look for additional fermionic symmetry generators constructed from
$\tilde{Q}$. Using the Galilean boost we find
$$
[\tilde{Q},b^j]:=\tilde{Z}^j,\qquad
\tilde{Z}^j=\frac{\epsilon_{jk}}{\sqrt{2m}}\pmatrix{m\sigma^k&0\cr-\partial_k&-m\sigma^k}.
$$
It is important to note, that
\be
\tilde{Z}^j=-\epsilon_{jk}Z^k.
\label{ZZtilde}
\ee
Then, the (anti)commutators among the
$\tilde{Z}^j$-s are obtained simply from those among the $Z^j$-s:
\be
\{\tilde{Z}^j,\tilde{Z}^k\}=m\delta^{jk}\mathrm{1}_4,
\label{Zt1}
\ee
Furthermore a direct computation shows
\be
\{\tilde{Q},\tilde{Z}^j\}=i\partial_j\mathrm{1}_4.
\label{Zt2}
\ee
Using the generator of expansion one finds
$$
[\tilde{Q},K]:=\tilde{S},\qquad
\tilde{S}=\frac{1}{\sqrt{2m}}
\pmatrix{-i\epsilon_{kl}\sigma^k(t\partial_l-imx_l)&0\cr-i\sigma^3-(x_1\partial_2-x_2\partial_1)&i\epsilon_{kl}\sigma^k(t\partial_l-imx_l)},
$$
satisfying also $\{\tilde{S},\nablaslash\}=0$ and 
$[d,\tilde{S}]=\tilde{S}$. Computing the square of this operator and exploiting the identification
(\ref{thid}) we find
$$
\tilde{S}\tilde{S}=iK+\frac{t}{2m}\gamma^t\nablaslash.
$$
Since the second term gives zero on solutions of LL we can say that
\be
\tilde{S}\tilde{S}=iK \label{StSt}
\ee
holds weakly in this case. One finds also
$$
[\tilde{Z}^j,b^k]=0,\qquad [\tilde{S},b^j]=0,\qquad [\tilde{Z}^j,K]=0,
\qquad [\tilde{S},K]=0.
$$

Thus we see that both the
$(\Lambda, Z^j, \hat{S})$ set and the $(\tilde{Q}, \tilde{Z}^j, \tilde{S})$ one
give a fermionic extension of the bosonic symmetry: the square of the
supercharges (respectively of the conformal supercharges) 
gives the (appropriate form of) Hamiltonian (respectively the operator of 
expansion), the two sets of $Z$-s form two fermionic Heisenberg algebras, and
the anticommutator of the supercharges with the corresponding $Z$-s give spatial translations. The still missing anticommutators within each set are the
following:
\be
\{\hat{S},Z^j\}=ib^j,\quad\{\Lambda,\hat{S}\}=i d,\qquad\{\tilde{S},\tilde{Z}^j\}=ib^j,\quad
\{\tilde{Q},\tilde{S}\}=id+\frac{1}{2m}\gamma^t\nablaslash .
\label{hianyzok}
\ee
(In the last equality the second term vanishes again on solutions of LLE, thus
``weakly'' it is absent). Combining (\ref{hianyzok}) with the previous
anticommutators
(\ref{lambdakuanti}, \ref{thid}, \ref{Z1}, \ref{Z2}, \ref{SS}, \ref{Zt1},
\ref{Zt2}, \ref{StSt}) one
can conclude, that extending the bosonic algebra (time and space translations,
rotation, Galilean boosts, dilatation and expansion) with either the
$(\Lambda,Z^j,\hat{S})$ or the $(\tilde{Q},\tilde{Z}^j,\tilde{S})$ sets results
in two closing super (i.e. $Z_2$ graded) algebras, the structure of which is an
$N=1$ (super) extension of the 2dim. Schr\"odinger symmetry
\cite{Duval:1993hs} \cite{Duval:1995dq}. 
The only difference
between the two cases is that for the first extension the algebra closes
without using the LLE, while in the second case one has to use it.  

The interesting question is whether one can use the two fermionic sets
{\bf simultaneously} to extend the bosonic symmetry. To decide this one has to
compute the various anticommutators between the elements of the
{\bf different} sets
and check whether the algebra closes. Some of these anticommutators are easy to
compute exploiting $\{\Lambda,\tilde{Q}\}=0$ and the identification between the
$Z^j$-s and $\tilde{Z}^j$-s, and the results stay within the algebra. However
$\{\Lambda,\tilde{S}\}$ (which also equals $-\{\tilde{Q},\hat{S}\}$) requires
a direct computation:
$$
\{\Lambda,\tilde{S}\}=iJ+Y\qquad
Y=-
\frac{i\alpha}{\sqrt{2m}}\pmatrix{im\sigma^3&0\cr\epsilon_{pq}\sigma^p\partial_q&im\sigma^3}=\frac{-i}{2m}\pmatrix{im\sigma^3&0\cr\epsilon_{pq}\sigma^p\partial_q&im\sigma^3}
$$
The first term on the r.h.s. is proportional to rotation, thus it is there in
the bosonic symmetry algebra, but the second term is a new (bosonic) one. Thus
we have to check whether the 
symmetry algebra
is closed even after including this new
element. 
One finds, remarkably, that $Y$ {\bf commutes with all generators of bosonic
  symmetry}:
\be
[{\cal B},Y]=0,\qquad{\cal B}=i\partial_t,\ i\partial_j,\ b^j,\ d,\ K,\ J.
\label{bx}
\ee
Furthermore,
\be
[\Lambda,Y]=i\tilde{Q},
\label{lambdax}
\ee
and algebraically
$$
[\tilde{Q},Y]=\frac{-i}{\sqrt{2m}}\pmatrix{-i\sigma^l\partial_l&0\cr\frac{i}{m}\partial_k\partial_k&i\sigma^l\partial_l}.
$$
However, using the identification (\ref{thid}) $\partial_k\partial_k=-2mi\partial_l$ we can write
$$
[\tilde{Q},Y]=\frac{-i}{\sqrt{2m}}\pmatrix{0&2mi\cr\partial_t&0}+
\frac{-i}{\sqrt{2m}}\pmatrix{-i\sigma^l\partial_l&-2mi\cr\partial_t&i\sigma^l\partial_l}=-i\Lambda -\frac{i}{\sqrt{2m}}\nablaslash.
$$
Since the second term vanishes on solutions of LLE, we can write, that weakly
\be
[\tilde{Q},Y]=-i\Lambda.\label{Qkigyox}
\ee
 The commutators between $Y$ and the additional 
fermionic generators are obtained from the definitions of these operators and
(\ref{bx}, \ref{lambdax}, \ref{Qkigyox}):
$$
[Z^j,Y]=i\tilde{Z}^j,\qquad [\hat{S},Y]=i\tilde{S},\qquad [\tilde{S},Y]=-i\hat{S}.
$$
Thus we conclude, that one can safely include $Y$ in the algebra; in fact its
properties (commuting with all bosonic generators and transforming the two
supercharges into each other) remind that bosonic $U(1)$ present 
in $N=2$ superconformal algebra.

Thus, looking at the (anti)commutators
(\ref{lambdakuanti}, \ref{thid}, \ref{Z1}, \ref{Z2}, \ref{SS}, \ref{Zt1},
\ref{Zt2}, \ref{StSt}, \ref{hianyzok}, \ref{bx}, \ref{lambdax}, \ref{Qkigyox}), 
we see that the generators of the Schr\"odinger symmetry
accompanied by $Y$ form the bosonic generators
$$
{\cal B}=i\partial_t,\ i\partial_j,\ b^j,\ d,\ K,\ J,\ Y,
$$
while the 
$(\Lambda,Z^j,\hat{S})$ and the $(\tilde{Q},\tilde{Z}^j,\tilde{S})$ sets the
fermionic ones
$$
{\cal F}=\Lambda,\ Z^j,\ \hat{S},\ \tilde{Q},\ \tilde{Z}^j,\ \tilde{S}
$$
of an $N=2$ superalgebra. Taking into account the central element and the
relation between $Z^j$ and $\tilde{Z}^k$, (\ref{ZZtilde}), we conclude, that
this algebra is the special two dimensional one of
\cite{Duval:1993hs} \cite{Duval:1995dq}.  
We note that all bosonic generators commute
with $\Gamma$, but the fermionic ones anticommute with it
$$
[{\cal B},\Gamma]=0,\qquad\quad \{{\cal F},\Gamma\}=0,
$$
thus we need both chirality spinors to represent this algebra. Also  
it is important to emphasize that to show the closure
of this algebra one has to use the LL equations, i.e. the algebra closes
``weakly'', on the solution space of LLEs. This manifests itself in the
appearance of weakly vanishing terms in various (anti)commutators, and we show
in Appendix A that these terms preserve weakly the generalized (``graded'')
Jacobi identity.

\subsection{Remarks, discussion}

Perhaps it is enlightening to point out that although this
$N=2$ extension of the Schr\"odinger symmetry is the same as the one found with
respect of the planar Pauli equation \cite{Duval:1995dq},
in fact it is generated (in part at least)
by operators which are related to the ones in the
Pauli equation problem in a rather tricky way. What I mean is that, obviously, $\tilde{Q}$ is the
4d version of the ``twisted'' $Q_{\rm 2d}$ used in the Pauli problem, however
$\Lambda$ seems to be related to $Q_{\rm 2d}$ in a surprising way.
In fact the straightforward 4d
generalization of  $Q_{\rm 2d}$ 
$$
Q=\frac{1}{\sqrt{2m}}\pmatrix{-i\sigma^k\partial_k&0\cr 0&-i\sigma^k\partial_k}
$$
(that also squares to $H$: $QQ=H$) 
{\bf commutes} (rather than anticommutes)
with the Dirac operator, $[Q,\nablaslash ]=0$. Therefore, although it 
generates a symmetry of the LLE, the symmetry it generates is a {\bf bosonic
  one}, and most likely 
$Q$ itself is (the light-like reduction of) one of the many
bosonic symmetries found in \cite{Durand:1988kd} for the free massless Dirac
equation in Minkowski space.
On the other hand algebraically
$$
Q=\Lambda +\frac{1}{\sqrt{2m}}\pmatrix{1&0\cr0&-1}\nablaslash ,
$$
which means, that weakly, on the solution space of the LLE, the two operators
coincide. 
(I
was unable to find a 4d generalization of $Q_{\rm 2d}$ that would anticommute with
the Dirac operator).

It would be interesting to see whether the $N=2$ extension found here 
can be given some
additional structure somewhat similarly to  \cite{Aizawa:2016vmo}, 
\cite{Aizawa:2016csp} \cite{Aizawa:2017ova}.

To emphasize that our findings depend crucially on working in $2+1$ dimensions
we note that in $d$ (space) $+\ 1$ (time) dimensions there is just one LLE
when $d$ is odd, while the number of LLE-s is two if $d$ is even. This can be
seen in our "light-like" Kaluza Klein framework in the following way
\cite{Palla:2020aa}: this
time we start with the free massless Dirac equation on $D=d+2$ dimensional
Minkowski space. The $2^{[D/2]}$ dimensional (Dirac) spinor representation is
an irreducible one when $D$ is odd (and in this case there is no chirality
matrix), while for even $D$ there is a chirality matrix and it splits the
Dirac spinor representation into two ($2^{[d/2]}$ dimensional) Weyl spinor
representations. Therefore when $D$ ($d$) is odd there is just one LLE after 
the Kaluza Klein reduction, while
when $D$ ($d$) is even, there are two LLE-s,   
since the chirality matrix preserves $\xi$, and it splits also the reduced Dirac
equation into two independent equations for the two Weyl spinors. 
 The appropriate generalization of $\Lambda$ works
for any $d$ independently whether $d$ is even or odd, and the corresponding
$N=1$ extension of the Schrödinger symmetry exists. However in the even case
$\Lambda$ 
anticommutes with the chirality matrix indicating it maps the two Weyl spinors
into
each other, therefore we need their direct sum to represent the extension, 
just like for $d=2$. To have "more" than $N=1$ SUSY we would 
need more
supercharges (like in $d=2$), but from the epsilon tensor, 
the gamma matrices and first derivatives
one can make a scalar only in two dimensions. Thus the $N=2$ extension exists
only in the planar case, in accord with \cite{Duval:1993hs}.

\section{Search for SUSY in the gauged Lévy-Leblond equations}

\subsection{The gauged Lévy-Leblond equations and its conformal symmetry}

 The {\bf gauged} LL equations are obtained by ``light-like'' reduction 
 from the gauged massless Dirac
 equation $\D\psi=0$, where $D_\mu=\nabla_\mu-iea_\mu$, with $a_\mu(x)$ being a
 $U(1)$ gauge field on $M_4$. Its field strength $f_{\mu\nu}=\partial_\mu a_\nu -
 \partial_\nu a_\mu$ satisfies the Bianchi identity $\partial_{[ \mu} f_{\nu\rho ]}
 =0$, (i.e. $2f=f_{\mu\nu}dx^\mu \wedge dx^\nu$ is a closed two form), but at this
 point  we assume no dynamical equation for it and treat the gauge
 field as an ``external'' one. Since we are concerned here with the light like
 reduction of the $4d$ Dirac equation we have to impose some condition on
 $f_{\mu\nu}$ ($a_\mu$) that guarantees the possibility of this. A useful
 condition is
 \be
 f_{\mu\nu}\xi^\nu =0, \label{fcond}
 \ee
 since through the Bianchi identity it guarantees, that $f_{\mu\nu}$ is a lift
 of a closed $2$ form $F_{\alpha\beta}$ ($\alpha\ \beta =t,1,2$) defined on
 $2+1$ dimensional non relativistic space time. Therefore without loss of
 generality $a_\mu$ can be chosen to be the lift of a vector potential
 $A_\alpha =(A_t,A_j)$ for $F_{\alpha\beta}$. Thus 
 effectively we get the gauged LLE from 
 (\ref{freeLL}) by the
 substitution $\partial_j\rightarrow D_j\equiv\partial_j-ieA_j$,
 $\partial_t\rightarrow D_t\equiv\partial_j-ieA_t$:
\begin{equation}
  \pmatrix{-i\sigma^jD_j&-2im\cr D_t&i\sigma^jD_j}
  \pmatrix{\Phi\cr\chi}=0\,.
  \label{gaugedLL}
 \end{equation}
  The crucial difference to the free case is that while the ordinary
 derivatives commute, the covariant ones do not, thus several new terms may
 appear. These terms show up already in the ``square'' of (\ref{gaugedLL}):
 $$
 \pmatrix{-D_j^2-e\sigma^3\epsilon_{kl}\partial_kA_l-2imD_t&0\cr
   -e\sigma^kF_{tk}&-D_j^2-e\sigma^3\epsilon_{kl}\partial_kA_l-2imD_t}
 \pmatrix{\Phi\cr\chi}=0\,.
 $$
 If we restrict our attention to static, purely magnetic gauge 
 fields 
 then $F_{tk}\equiv 0$, and $D_t=\partial_t$, and this equation can be written:
 $$
i\partial_t\pmatrix{\Phi\cr\chi}=H_e\pmatrix{\Phi\cr\chi}, \quad\qquad 
H_e=-\frac{1}{2m} \pmatrix{D_j^2+e\sigma^3\epsilon_{kl}\partial_kA_l&0\cr
   0&D_j^2+e\sigma^3\epsilon_{kl}\partial_kA_l}.
$$

In  \cite{Duval:1995fa} we investigated the symmetries of the gauged LLE,
(\ref{gaugedLL}), and showed, that the infinitesimal $\xi$ preserving
conformal transformations satisfy
\be
   [{\cal B},\D]=-ie\gamma^\mu(L_{X_{\cal B}}a)_\mu
   -\frac{1}{4}\nabla_\mu X_{\cal B}^\mu \D ,
   \label{bbsym}
   \ee
   where $(L_{X_{\cal B}}a)_\mu$ is the Lie derivative of the gauge field with
   respect to $X_{\cal B}^\mu\partial_\mu$. 
   Note that
   - because of the terms with the Lie derivative - this equation is more
   complicated then (\ref{symeqb}) in the case of the free LLE. Therefore if
   $\psi$ solves the gauged LLE, (\ref{gaugedLL}), then ${\cal B}\psi$ solves
   rather
   $$
   \D({\cal B}\psi)-ie\gamma^\mu(L_{X_{\cal B}}a)_\mu\psi=0,
   $$
   i.e. the transformed form of the gauged LLE. For later reference we list here
   the explicit form of the commutators between the various ${\cal B}$-s and
   $\D$ as they follow from eq.(\ref{bbsym}),  
   when the gauge field is static and purely magnetic (i.e. when $A_t=0$,
$\partial_tA_j=0$):
 \be
    [i\partial_j,\D]=-ie\gamma^k(i\partial_jA_k),\qquad\quad [i\partial_t,\D]=0,
    \quad\qquad
    [J,\D]=-ie\gamma^k(\epsilon_{lj}x^l\partial_jA_k+\epsilon_{kj}A_j),
    \label{com1}
    \ee
\be    
   [d,\D]=-ie\gamma^k(x^j\partial_jA_k+A_k)-\D,\quad\qquad 
   [b^j,\D]=-ie\bigl(\gamma^k(t\partial_jA_k)+\gamma^tA_j\bigr), \label{com2}
   \ee
\be
   [K,\D]=-ie\bigl(\gamma^kt(x^j\partial_jA_k+A_k)+\gamma^tA_mx^m\bigr)-t\D .\label{com3}
   \ee

 \subsection{The fermionic extension in case of the gauged LLE}

 Next we look whether the supercharges and the associated fermionic extensions
 found in the case of the free LLE work also for the gauged LLE. We start with
 $\Lambda$, eq.(\ref{alfabeta}), and find that for a static, purely
 magnetic gauge field it anticommutes with $\D$:
 \be
 \{\Lambda,\D\}=0,\qquad {\rm if}\quad A_t=0,\quad {\rm and}\quad
 \partial_tA_j=0.\label{lambdaD}
 \ee
 This is a good sign, however it is not enough as in the case of the free LLE,
 and to progress we have to check whether the modifications in eq.(\ref{com1}-
 \ref{com3}) represented by the terms with the Lie derivatives are consistent
 with the algebra of the extension $(\Lambda,\ Z^j,\ \hat{S})$.

 To start to investigate this we use two identities, both of which are
 obtained by simple algebra exploiting eq.(\ref{lambdaD}); a ``bosonic'' one
 \be
 \{[\Lambda,{\cal B}],\D\}=\{\Lambda,[{\cal B},\D]\},\label{bid}
 \ee
valid for any bosonic generator ${\cal B}$, and a ``fermionic'' one
 \be
    [\Lambda,\{{\cal F},\D\}]=[\{\Lambda,{\cal F}\},\D], \label{fid}
    \ee
    valid for any fermionic generators.

    $\Lambda$ commutes with translations (${\cal B}=i\partial_j$) and rotation
    (${\cal B}=J$) thus when (\ref{bid}) applied in these cases the l.h.s.
 vanishes. Thus we have to check whether $\Lambda$ indeed anticommutes with     
 the r.h.s. of the first and third expressions in (\ref{com1}). One can prove
 in general, that
 \be
 \{\Lambda,W\}=0,\qquad{\rm for}\quad W=-ie\gamma^kW_k.\quad
 {\rm provided}\quad\partial_tW_k=0,\label{LW}
 \ee
 and the explicit expressions of $[i\partial_j,\D]$ and $[J,\D]$
 are precisely of this
 form. Thus we conclude that $[\Lambda,i\partial_j]=0$ and $[\Lambda,J]=0$ are
 consistent with (\ref{com1}).

 We also notice, that applying (\ref{bid}) for ${\cal B}=d$ we get zero on the
 r.h.s. when using eq.(\ref{com2}) 
 as a consequence of (\ref{lambdaD}) and (\ref{LW}). However this is
 consistent, since on the l.h.s. $[\Lambda,d]=\Lambda$ and $\{\Lambda,\D\}=0$.

 Applying (\ref{bid}) for ${\cal B}=b^j$ (${\cal B}=K$) determines the
 anticommutators of $Z^j$ ($\hat{S}$) and $\D$ explicitly:
  \be
  \{Z^j,\D\}={\cal N}^j,
  \qquad{\cal N}^j=
  \frac{e}{\sqrt{2m}}\pmatrix{2mA_j&0\cr-\sigma^k\partial_jA_k&2mA_j},
 \label{ZD}
 \ee
 and
 \be
 \{\hat{S},\D\}={\cal L}-\frac{1}{\sqrt{2m}}\gamma^t\D,\quad\qquad
 {\cal L}=\frac{-ie}{\sqrt{2m}}\pmatrix{2miA_mx^m&0\cr-i\sigma^k{\cal M}_k&2miA_mx^m},
       \label{SD}
       \ee
 with ${\cal M}_k=x^j\partial_jA_k+A_k$. The ${\cal N}^j$ and ${\cal L}$ are  
the equivalents of the Lie derivative terms in (\ref{bbsym}): if $\psi$ 
solves $\D\psi=0$, then, on the basis of (\ref{ZD}, \ref{SD}) $Z^j\psi$ and
$\hat{S}\psi$ solve the transformed equations
$$
\D(Z^j\psi)-{\cal N}^j\psi=0,\qquad\quad\D(\hat{S}\psi)-{\cal L}\psi=0.
$$
The consistency of these terms with (\ref{com1}-\ref{com3}) is investigated in
 Appendix B.

 We apply the fermionic identity, eq.(\ref{fid}), first for ${\cal F}=Z^j$. The
 l.h.s. can be written
 $$
 [\Lambda,\{Z^j,\D\}]=[\Lambda,\{\Lambda,[b^j,\D]\}]=[\Lambda\Lambda,[b^j,\D]]
 =[i\partial_t,[b^j,\D]]=i\partial_t([b^j,\D])=[i\partial_j,\D],
 $$
 where in the first equality the definition of $Z^j$ and 
 (\ref{bid}) is used, the second equality is simple
 algebra, the third equality uses the square of $\Lambda$ while in the last
 equality we used the explicit form of $[b^j,\D]$, (\ref{com2}), and the fact
 that $\partial_tA_j=0$. Thus we conclude that (\ref{com1}-\ref{com3}) are also
 consistent with $\{\Lambda,Z^j\}=i\partial_j\mathrm{1}_4$.

 In a similar way applying (\ref{fid}) for ${\cal F}=\hat{S}$ we get
$$
 [\Lambda,\{\hat{S},\D\}]=[\Lambda,\{\Lambda,[K,\D]\}]=[\Lambda\Lambda,[K,\D]]
 =[i\partial_t,[K,\D]]=i\partial_t([K,\D])=[id,\D],
 $$
 where the last equality is based on (\ref{com3}) with $\partial_tA_m=0$ and
 (\ref{com2}).  Thus we conclude that (\ref{com1}-\ref{com3}) are also
 consistent with $\{\Lambda,\hat{S}\}=id$.

 Unfortunately we found no general framework to check the consistency of the
 rest of the fermionic anticommutation relations and eq.(\ref{com1}-
 \ref{com3}) thus we have to resort to a case by case analysis. We collect
 some of these not very illuminating computations in Appendix B and here
 merely state that the outcome is positive: the fermionic algebra generated by
 $(\Lambda,\ Z^j,\ \hat{S})$ in case of the free LLE is consistent with
 eq.(\ref{com1}-\ref{com3}), thus the whole $N=1$ super extension of the
 Schrödinger symmetry works also for the gauged LLE at least when the
 external gauge field is static and purely magnetic.

 The situation of the other supercharge, $\tilde{Q}$, and the associated
 extension $(\tilde{Q},\ \tilde{Z}^j,\ \tilde{S})$ is different.
 $\tilde{Q}$ has several properties, that make it a potential supercharge
  for the free LLE: e.g. 
 it anticommutes with $\nablaslash$ and its square is $H$. None of these
 survive for the gauged LLE, no matter what kind of gauge field we have:  
 $\{\tilde{Q},\D\}\neq 0$, and $\tilde{Q}\tilde{Q}=H\neq H_e$. Interestingly 
 one can define an ``external field'' version of $\tilde{Q}$
 $$
 \tilde{Q}_e=\frac{1}{\sqrt{2m}}
\pmatrix{-i\epsilon_{kl}\sigma^kD_l&0\cr 0&i\epsilon_{kl}\sigma^kD_l},
$$
that squares to $H_e$: $H_e=\tilde{Q}_e\tilde{Q}_e$. Furthermore it 
anticommutes with $\D$ when the gauge field is static and purely magnetic 
$$
\{\tilde{Q}_e,\D\}=0,\qquad {\rm if}\quad A_t=0,\quad {\rm and}\quad
 \partial_tA_j=0,
$$
thus in this case 
it generates a symmetry of the gauged LLE, (\ref{gaugedLL}). However it
can not play the role of a supercharge in a fermionic extension of the
Schrödinger symmetry, since it does not commute with
translations
$$
[i\partial_j,\tilde{Q}_e]=\frac{e}{\sqrt{2m}}\gamma^k\epsilon_{kl}\partial_jA_l
\neq 0,
$$
and the algebra it generates would depend on the external field.

\section{SUSY in the coupled  Lévy-Leblond and Chern-Simons equations}

\subsection{The gauged Lévy-Leblond equations coupled to Chern-Simons
  theory}

Next we investigate the fermionic extension of the Schrödinger symmetry when
the dynamics of the gauge field appearing in (\ref{gaugedLL}) is determined by
the Chern-Simons (CS) field equations.

In \cite{Duval:1994pw} 
it is shown that on a general Bargman space the 4d form of the
CS equations is the field current identity
\be
f_{\mu\nu}=\frac{e}{\kappa}\sqrt{-g}\epsilon_{\mu\nu\rho\sigma}\xi^\rho j^\sigma,
\qquad f_{\mu\nu}=\partial_\mu a_\nu -\partial_\nu a_\mu, \label{CSE}
\ee
where $\kappa$ is the CS coupling and $j^\mu$ is some 4d current.
Please note, that (\ref{CSE}) implies (\ref{fcond}), thus $f_{\mu\nu}$ is a
lift of a closed $F_{\alpha\beta}$. Furthermore in \cite{Duval:1994pw} 
we showed that
$j^\mu$ projects to a 3 current $J^\alpha=(\rho,J^k)$ ($\alpha=t,1,2$) and
(\ref{CSE}) descends in the lightlike reduction as
$$
F_{\alpha\beta}=-\frac{e}{\kappa}\sqrt{-g}\varepsilon_{\alpha\beta\gamma}J^\gamma .
$$ 
On our Minkowski space with metric $ds^2=\sum (dx^i)^2 +dtds$ this can be written
as
\be
B\equiv\epsilon_{ij}\partial_iA_j=-\frac{e}{\kappa}\rho,\qquad {\rm and}
\qquad E^j\equiv F_{jt}=\frac{e}{\kappa}\epsilon_{jk}J^k,
\label{CSEl}
\ee
which are indeed the CS field equations in \cite{Jackiw:1990cs}. Since
$F_{\alpha\beta}$ is closed, ($\partial_tB+\epsilon_{ij}\partial_iE^j=0$),
$\rho\equiv J^t$ and $J^j$ must satisfy the reduced conservation equation
\be
\partial_tJ^t +\partial_jJ^j =0.
\label{Rc}
\ee

Now we couple the CS equations to the LLE by identifying the CS current in
(\ref{CSE}) with the natural (conserved) current associated to the gauged
LLE. This current is made of the spinor fields and to construct it we need
the Dirac adjoint $\bar{\psi}=\psi^\dagger G$, where $G$ is determined by
the requirements $\bar{\gamma}_\mu:=G^{-1}\gamma_\mu^\dagger G=\gamma_\mu$ and
$G^\dagger =G$. One can show that $\D\bar{\psi}=0$ whenever $\D\psi=0$, thus,
as a consequence
\be
\nabla_\mu(\bar{\psi}\gamma^\mu\psi)=0.
\label{curc}
\ee
Therefore we can identify $j^\mu$ in (\ref{CSE}) and
$\bar{\psi}\gamma^\mu\psi$:
$$
j^\mu=\bar{\psi}\gamma^\mu\psi .
$$
With our Dirac matrices (\ref{gm}) $G$ turns out to be $G=\pmatrix{0&1\cr1&0}$,
and as a consequence of the particular form of $\gamma^t$
\be
\rho\equiv J^t=\vert\Phi\vert^2=\vert\phi_+\vert^2+\vert\phi_-\vert^2.
\label{rho}
\ee
The spatial components of the current are
\be
J^j=i(\Phi^\dagger \sigma^j\chi-\chi^\dagger\sigma^j\Phi).
\label{jj}
\ee
Note that the reduced conservation equation (\ref{Rc}) for this $\rho$ and $J^j$
follows from (\ref{curc}) since $\bar\psi\gamma^s\psi\sim \chi^\dagger\chi$ is
independent of $s$.

Thus the system of coupled LL and CS equations is given by (\ref{gaugedLL}),
(\ref{CSEl}) and (\ref{rho}-\ref{jj}). In  \cite{Duval:1995fa} 
we showed that the $\xi$
preserving conformal transformations act as symmetries on the solutions of this
system. Here we are interested whether the $N=1$ superextension of these
symmetries, represented by $(\Lambda,\ Z^j,\ \hat{S})$, is also a symmetry of
the coupled system.

\subsection{Fermionic extension for the coupled LL and CS equations}

In the previous section we showed that the algebra generated by
$(\Lambda,\ Z^j,\ \hat{S})$ is a symmetry of the gauged LLE if the gauge field
satisfies
$$
A_t\equiv 0,\qquad\quad {\rm and}\qquad\quad \partial_tA_j=0.
$$
While for the gauged LLE alone these conditions are relatively harmless, here,
in the coupled system, they impose some non trivial restrictions, and - as we
argue below - require that we consider only static solutions of
(\ref{gaugedLL}) with {\bf definite chirality} spinors.

These conditions imply that $F_{jt}$ vanishes, thus, because of (\ref{CSEl}),
\be
J^j=0,\qquad\quad{\rm and\ also}\qquad\quad \partial_t\rho=0,
\label{jrho}
\ee
must also hold. In light of (\ref{rho}) and (\ref{jj}) these requirements are
satisfied if we look for solutions of (\ref{gaugedLL}) with\footnote{The other
  generic solution of (\ref{jrho}), when $\Phi\equiv0$, leads to vanishing
  $\chi$ and pure gauge $A_j$, thus is not interesting.}    
\be
\chi\equiv 0,\qquad\quad \partial_t\Phi=0,
\label{chiphi}
\ee
i.e. if (\ref{gaugedLL}) simplifies to
$$
-i\sigma^jD_j\Phi=0,\qquad\quad \partial_t\Phi=0.
$$
In terms of the spinors with definite chirality these equations mean, that
$\chi_+=0$, $\chi_-=0$ and their nonvanishing components are static and satisfy
\be
(D_1+iD_2)\phi_+=0,\qquad\quad (D_1-iD_2)\phi_-=0,
\label{sd1}
\ee
respectively. Although the equations for $\phi_+$ and $\phi_-$ look
independent, they contain the same gauge field $A_j$ and the first eq. in
(\ref{CSEl}) couples them as
\be
\epsilon_{ij}\partial_iA_j=-\frac{e}{\kappa}(\vert\phi_+\vert^2+
\vert\phi_-\vert^2).
\label{sd2}
\ee
One can show that (\ref{sd1}-\ref{sd2}) admit normalizable solutions when only
one of the $\phi_\pm$-s is different from zero. If $\kappa <0$ then the
normalizable solution exists for $\phi_+$ while for $\kappa >0$ it exists for
$\phi_-$. (See Appendix C).

If under an infinitezimal fermionic transformation the spinors change as
$$
\pmatrix{\Phi\cr\chi}\rightarrow\pmatrix{\Phi\cr\chi}+
\epsilon\pmatrix{\tilde{\Phi}\cr\tilde{\chi}},
\qquad \pmatrix{\tilde{\Phi}\cr\tilde{\chi}}={\cal F}\pmatrix{\Phi\cr\chi},
\qquad {\cal F}=\Lambda,\,Z^j,\,\hat{S},
$$
then, in general, $\rho$ and $J^j$ also change
\be
\rho\rightarrow\rho+\delta\rho=\rho+\epsilon^*\tilde{\Phi}^\dagger\Phi+
\epsilon\Phi^\dagger\tilde{\Phi},
\label{drho}
\ee
$$
J^j\rightarrow J^j+\delta J^j=J^j+
i\bigl(\epsilon\Phi^\dagger\sigma^j\tilde{\chi}+
\epsilon^*\tilde{\Phi}^\dagger\sigma^j\chi-
\epsilon\chi^\dagger\sigma^j\tilde{\Phi}-
\epsilon^*\tilde{\chi}^\dagger\sigma^j\Phi\bigr).
$$
Note that for solutions of our interest, i.e. when (\ref{chiphi})
holds, $\delta J^j$ simplifies to
\be
\delta J^j=i\bigl(\epsilon\Phi^\dagger\sigma^j\tilde{\chi}-
\epsilon^*\tilde{\chi}^\dagger\sigma^j\Phi\bigr).
\label{dj}
\ee
The coupled LL and CS equations admit the fermionic symmetries, since for
any ${\cal F}$ the $\delta\rho$ and $\delta J^j$ in (\ref{drho}), (\ref{dj}) 
vanish (thus the CS equations
preserve their form). Fortunately to show this
there is no need to determine $\tilde{\Phi}$ and $\tilde{\chi}$ explicitly
for the various ${\cal F}$-s. Indeed recalling that $\{\Gamma,{\cal F}\}=0$
for all ${\cal F}$, we see that when we start with a positive chirality
solution then its
fermionic transform is of negative chirality
$$\pmatrix{\Phi\cr\chi}=\pmatrix{\phi_+\cr 0\cr 0\cr 0},\qquad\quad  
\pmatrix{\tilde{\Phi}\cr\tilde{\chi}}={\cal F}\pmatrix{\Phi\cr\chi}=
\pmatrix{0\cr\tilde{\phi}_-\cr\tilde{\chi}_-\cr0},
$$
(for some $\tilde{\phi}_-$, $\tilde{\chi}_-$) and vice versa
$$
{\cal F}\pmatrix{0\cr\phi_-\cr 0\cr 0}=
\pmatrix{\tilde{\phi}_+\cr 0\cr 0\cr\tilde{\chi}_+}.
$$
Since the various terms
determining  $\delta\rho$ and $\delta J^j$  in (\ref{drho}), (\ref{dj}) 
couple only same chirality spinors
$$
\Phi^\dagger\tilde{\Phi}=\phi_+^\dagger\tilde{\phi}_++\phi_-^\dagger\tilde{\phi}_-,
\qquad\quad\Phi^\dagger\sigma^1\tilde{\chi}=
\phi_+^\dagger\tilde{\chi}_++\phi_-^\dagger\tilde{\chi}_-,
\qquad\quad\Phi^\dagger\sigma^2\tilde{\chi}=i(
\phi_+^\dagger\tilde{\chi}_+-\phi_-^\dagger\tilde{\chi}_-),
$$
it is obvious, that they give zero, when evaluated for the definite
chirality solutions and their fermionic transforms. 

\section{Conclusions}

In the central part of this paper we present an $N=2$ extension of Schrödinger
symmetry $sch(2)$ \cite{Duval:1993hs} \cite{Duval:1995dq} for the free planar
LLE. This extension is built in terms of operators that anticommute with the
LL differential operator. The construction is based on several special
properties of the planar problem e.g. on the existence of two LLE-s, the
solutions of which span two representations of $sch(2)$, describing non
relativistic spin $(1/2)$ and spin $(-1/2)$ particles respectively
\cite{Duval:1995fa}. It is also a
special property of two spatial dimensions, that we can find two supercharges;
one, ($\Lambda$), with square $i\partial_t$, and another one, ($\tilde{Q}$),
which squares to the free Pauli Hamiltonian. Both of these operators map the
two non relativistic spinor representations into each other, thus we need both
representations to construct the $N=2$ extension. The  $N=2$ extension
requires the
equality of $\Lambda\Lambda$ and $\tilde{Q}\tilde{Q}$, we achieve this by
identifying - on the solution manifold of the free LLE - time translation and
the Pauli Hamiltonian. (This identification is made possible by the fact that
all solutions of any of the two LLE-s satisfy the Pauli equation). As a
consequence of this identification some of the (anti)commutators of the
extended $N=2$ algebra contain weakly vanishing terms, i.e. terms, which are
non zero algebraically, but vanish on solutions of the LLE. Thus we can say
that the $N=2$ algebra closes weakly, on solutions of LLE. 
We also show that the weakly vanishing
terms do not spoil the generalized Jacobi identity of the algebra.

Next we show that when the LLE is coupled to an external gauge field, the
$N=1$ part of the previous extension generated by $\Lambda$ persists as 
symmetry, at least when 
the external gauge field is static and purely magnetic. Since $sch(2)$ acts on
the gauged LLE in a more complicated way than on the free one, the major task
here is to show that the terms with the Lie derivatives of the gauge field
(which form that complicating difference to the free case) are 
consistent with the $N=1$ extension. 

Finally we show that the same $N=1$ extension persists as symmetry when the
dynamics of the gauge field is described by the Chern Simons field equations,
i.e. when we couple LLE to Chern Simons theory. This conclusion is based on
the observation that through the Chern Simons equations a static and purely
magnetic gauge field leads to static solutions of the LLE with definite
chirality spinors only. 

\section*{Acknowledgments}

I intended to present the results of this paper on the conference celebrating
the retirement of prof. Peter Horvathy, but this conference has been
postponed because of the pandemic. I thank Peter for 
his enthusiastic interest in these matters and for his remarks.

\appendix

\section{Check of the generalized Jacobi identity}

In this appendix we show that the various weakly vanishing terms appearing in
some of the (anti)commutators of the $N=2$ algebra keep the weak form of the
generalized Jacobi identity as they generate weakly vanishing terms only.

We have the following types of (anti)commutators containing weakly vanishing
terms
$$
[H,K]=id+\frac{1}{2m}\gamma^t\nablaslash,\qquad\quad 
\{\tilde{S},\tilde{S}\}=2iK+\frac{t}{m}\gamma^t\nablaslash ,
$$
$$
[\tilde{Q},Y]=-i\Lambda-\frac{i}{\sqrt{2m}}\nablaslash,\qquad\quad 
\{\tilde{Q},\tilde{S}\}=id+\frac{1}{2m}\gamma^t\nablaslash .
$$
The commutator on the l.h.s. of the third expression is a fermionic operator,
thus in the generalized Jacobi identity its commutators with the bosonic
generators ${\cal B}$ and anticommutators with the fermionic generators
${\cal F}$ appear. The commutators on the l.h.s. of the other three expressions
are bosonic operators, thus in the generalized Jacobi identity only their
commutators appear with both the  ${\cal B}$-s and the ${\cal F}$-s. Thus we
have to show that in these commutators/anticommutators the weakly vanishing
terms produce further weakly vanishing terms only.

The case of the third commutator is simple:
$$
[{\cal B},[\tilde{Q},Y]]=-i[{\cal B},\Lambda]-
\frac{i}{\sqrt{2m}}[{\cal B},\nablaslash ],\quad 
\{{\cal F},[\tilde{Q},Y]\}=-i\{{\cal F},\Lambda\}-
\frac{i}{\sqrt{2m}}\{{\cal F},\nablaslash\},
$$
since the symmetry equations (\ref{symeqb}) (\ref{symeq}) with the known
$\Sigma_{\cal B}$ and $\Sigma_{\cal F}=0$ 
show that the terms generated are indeed weakly vanishing.

Next we consider the (anti)commutators containing the weakly vanishing term
$\frac{1}{2m}\gamma^t\nablaslash$. In the above mentioned 
commutator with bosonic generators this term generates 
$$
[{\cal B},\frac{1}{2m}\gamma^t\nablaslash]=
\frac{1}{2m}\bigl([{\cal B},\gamma^t]\nablaslash+
\gamma^t[{\cal B},\nablaslash]\bigr)=
\frac{1}{2m}\bigl([{\cal B},\gamma^t]+
\gamma^t\Sigma_{\cal B}\bigr)\nablaslash .
$$
We are not yet ready, since we have to show that the term multiplying
$\nablaslash$ contains no derivatives. However 
since $[{\cal B},\gamma^t]+\gamma^t\Sigma_{\cal B}$ vanishes trivially
for ${\cal B}=i\partial_j,i\partial_t,J,b^j, Y$ (since both
$[{\cal B},\gamma^t]$ and $\Sigma_{\cal B}$ vanish) 
and non trivially for
${\cal B}=d$ ($[d,\gamma^t]=\gamma^t$ and $\Sigma_d=-1$) and ${\cal B}=K$
($[K,\gamma^t]=t\gamma^t$ and $\Sigma_K=-t$) we see that all the terms
generated vanish identically. In the commutator with fermionic generators
this term produces 
$$
[{\cal F},\frac{1}{2m}\gamma^t\nablaslash]=
\frac{1}{2m}\bigl([{\cal F},\gamma^t]\nablaslash+
\gamma^t[{\cal F},\nablaslash]\bigr)=
\frac{1}{2m}\bigl([{\cal F},\gamma^t]+
2\gamma^t{\cal F}\bigr)\nablaslash ,
$$
where we used that $\{{\cal F},\nablaslash\}=0$. The fermionic generators
${\cal F}=\tilde{Q},\tilde{S},Z^j,\tilde{Z}^j$ have the generic form
$$
{\cal F}=\pmatrix{A&0\cr B&-A}
$$
for some $2\times 2$ $A$ and $B$, and for them
$$
[{\cal F},\gamma^t]=\pmatrix{0&0\cr-2A&0},\qquad\qquad
2\gamma^t{\cal F}=2\pmatrix{0&0\cr A&0},\qquad {\rm thus}\qquad
[{\cal F},\gamma^t]+2\gamma^t{\cal F}=0.
$$
This argument does not apply for $\Lambda$ and $\hat{S}$, but a direct
computation gives
$$
[\Lambda,\gamma^t]+2\gamma^t\Lambda=i\sqrt{2m}\mathrm{1}_4,\qquad\quad
[\hat{S},\gamma^t]+2\gamma^t\hat{S}=\sqrt{2m}\pmatrix{it&0\cr-2\sigma^jx_j&it},
$$
and, since they contain no derivatives, we see that the terms generated for
them are indeed weakly vanishing.

The terms generated by the weakly vanishing $\frac{t}{m}\gamma^t\nablaslash$
require a separate consideration, as some bosonic and fermionic generators
contain time derivatives, thus the previous results may not apply in this case.
In case of the commutators with the bosonic generators
$$
[{\cal B},\frac{t}{m}\gamma^t\nablaslash]=
\frac{1}{m}\bigl([{\cal B},t\gamma^t]+
t\gamma^t\Sigma_{\cal B}\bigr)\nablaslash ,
$$
the term multiplying $\nablaslash$ vanishes trivially for all ${\cal B}$
apart from ${\cal B}=d$ and ${\cal B}=K$. However a direct computation gives
$$
[d,t\gamma^t]+t\gamma^t\Sigma_d=2t\gamma^t-t\gamma^t=t\gamma^t,\qquad\quad 
[K,t\gamma^t]+t\gamma^t\Sigma_K=2t^2\gamma^t+t\gamma^t(-t)=t^2\gamma^t,
$$
showing that they are weakly vanishing. In case of the commutators
with the fermionic operators
$$
[{\cal F},\frac{t}{m}\gamma^t\nablaslash]=
\frac{1}{m}\bigl([{\cal F},t\gamma^t]+
2t\gamma^t{\cal F}\bigr)\nablaslash ,
$$
the term multiplying $\nablaslash$ vanishes for 
${\cal F}=\tilde{Q},\tilde{S},Z^j,\tilde{Z}^j$ in the same way as before, since
they contain no time derivatives. For $\Lambda$ and $\hat{S}$ we find
$$
[\Lambda,t\gamma^t]+2t\gamma^t\Lambda=it\sqrt{2m}\mathrm{1}_4,\qquad\quad 
[\hat{S},t\gamma^t]+2t\gamma^t\hat{S}=t\sqrt{2m}\pmatrix{it&0\cr-2\sigma^jx_j&it}.
$$
Thus we conclude that  also 
$\frac{t}{m}\gamma^t\nablaslash$ generates only weakly
vanishing terms.

 \section{Consistency of the Lie derivative terms and the fermionic extension}

 Here we collect some of the case by case checks we carried out to prove that
 the terms with the Lie derivatives in (\ref{com1}-\ref{com3}) are consistent
 with the fermionic extension $(\Lambda,\ Z^j,\ \hat{S})$. We are concerned
 here mainly with the anticommutators of $Z^j$ ($\hat{S}$) and $\D$, obtained
 in (\ref{ZD}, \ref{SD}).

  The first anticommutator we check is $\{Z^j,Z^k\}$; a simple
computation gives
$$
[\{Z^j,Z^k\},\D]=[Z^j,\{Z^k,\D\}]+[Z^k,\{Z^j,\D\}]=[Z^j,{\cal N}^k]+
[Z^k,{\cal N}^j] .
$$
    Substituting here the expression one obtains using (\ref{ZD})
    and the explicit form of $Z^j$ 
$$
[Z^j,{\cal N}^k]=e\pmatrix{0&0\cr\partial_jA_k-\partial_kA_j&0}
$$
gives $[\{Z^j,Z^k\},\D]=0$, which is consistent with (\ref{Z2}).

Next we check the consistency of $\hat{S}\hat{S}=iK$:
$$
[\hat{S}\hat{S},\D]=[\hat{S},\{\hat{S},\D\}]=[\hat{S},{\cal L}]+
\frac{-1}{\sqrt{2m}} [\hat{S},\gamma^t\D],
$$
where we used (\ref{SD}). One finds explicitly:
$$
[\hat{S},{\cal L}]=e\pmatrix{-i\sigma^kt{\cal M}_k&0\cr0&i\sigma^kt{\cal M}_k},
$$
and
$$
\frac{-1}{\sqrt{2m}} [\hat{S},\gamma^t\D]=
-it\D+e\pmatrix{0&0\cr A_lx^l&0}.
$$
Thus
$$
[\hat{S}\hat{S},\D]=i\Bigl(-ie\pmatrix{-i\sigma^kt{\cal M}_k&0\cr
  A_lx^l&i\sigma^kt{\cal M}_k}-t\D\Bigr)=i[K,\D],
$$
where, in the last equality, eq.(\ref{com3}) is used.

Finally we show the consistency of $[d,Z^j]=0$ with the Lie derivative terms and
with (\ref{ZD}), i.e. we check whether $\{[d,Z^j],\D\}$ vanishes. A simple
algebra gives
\be
\{[d,Z^j],\D\}=[d,\{Z^j,\D\}]-\{Z^j,[d,\D]\}.
\label{dzc}
\ee
A direct computation using (\ref{ZD}) yields
$$
[d,\{Z^j,\D\}]=\frac{e}{\sqrt{2m}}\pmatrix{2mx^k\partial_kA_j&0\cr-\sigma^k\partial_j(x^m\partial_mA_k)&2mx^k\partial_kA_j},
$$
while, on the basis of (\ref{com2}) one can write
$$
\{Z^j,[d,\D]\}=-ie\{Z^j,\gamma^k{\cal M}_k\}-\{Z^j,\D\}.
$$
We find explicitly:
$$
-ie\{Z^j,\gamma^k{\cal M}_k\}=\frac{e}{\sqrt{2m}}\pmatrix{2m(x^l\partial_lA_j+
  A_j)&0\cr-\sigma^k\partial_j(x^l\partial_lA_k+A_k)&2m(x^l\partial_lA_j+A_j)},
$$
and, using (\ref{ZD}), get eventually
$$
\{Z^j,[d,\D]\}=\frac{e}{\sqrt{2m}}\pmatrix{2m(x^l\partial_lA_j)&0\cr-\sigma^k\partial_j(x^l\partial_lA_k)&2m(x^l\partial_lA_j)}.
$$
Since this is identical to $[d,\{Z^j,\D\}]$ above, we see that 
$\{[d,Z^j],\D\}$ in (\ref{dzc}) vanishes indeed.

\section{Solutions of (\ref{sd1}-\ref{sd2})}

In this appendix we study the solutions of (\ref{sd1}-\ref{sd2}). First we
assume, that none of the $\phi_\pm$-s vanishes identically. In this case,
introducing their modulus and phase $\phi_\pm=(\rho_\pm)^{1/2}e^{i\alpha_\pm}$
(where $\rho_\pm\ge 0$, and $\alpha_\pm$ are real), one can express the gauge
field from both equations in (\ref{sd1}); we get
$$
A_i=\frac{1}{2e}\epsilon_{ij}\partial_j\ln\rho_++\frac{1}{e}\partial_i\alpha_+
$$
from the equation for $\phi_+$, while
$$
A_i=-\frac{1}{2e}\epsilon_{ij}\partial_j\ln\rho_-+\frac{1}{e}\partial_i\alpha_-
$$
from the equation for $\phi_-$. However the two gauge fields must be the same,
thus
$$
\rho_-=\frac{1}{\rho_+},\qquad {\rm and}\qquad \alpha_-=\alpha_+
$$
must hold. Substituting this common gauge field into (\ref{sd2}) yields
$$
\bigtriangleup\ln\rho_+=\frac{2e^2}{\kappa}\bigl(\rho_++\frac{1}{\rho_+}\bigr).
$$
The r.h.s. of this equation is invariant under
$\rho_+\rightarrow\frac{1}{\rho_+}$, while the l.h.s. changes sign. Therefore
both sides of the equation must vanish:
$$
\bigtriangleup\ln\rho_+=0,\qquad\rho_++\frac{1}{\rho_+}=0.
$$
Since the second condition does not allow any real solutions, we conclude that
there is no solution of (\ref{sd1}-\ref{sd2}) with both $\phi_\pm$ non
vanishing.

On the other hand, if we assume, that only one of the $\phi_\pm$-s is different
from zero, then there is only one relevant equation in (\ref{sd1}) and from
this - after going into the $\phi_\epsilon=(\rho_\epsilon)^{1/2}$ gauge - 
we obtain one of the above expressions  for the
gauge field (with $\alpha_\epsilon\equiv 0$). Using this in
(\ref{sd2}) leads to the Liouville equation in both cases 
$$
\bigtriangleup\ln\rho_\pm=\pm\frac{2e^2}{\kappa}\rho_\pm.
$$
A normalizable solution is obtained for $\phi_+$ when $\kappa$ is negative,
while for $\phi_-$ when $\kappa$ is positive.

\bibliographystyle{unsrt}

\end{document}